\documentclass[aps,prd,onecolumn,groupedaddress,nofootinbib,showpacs,longbibliography,showkeys,amssymb]{revtex4-1}
\usepackage[colorlinks,citecolor=blue,urlcolor=blue,linkcolor=blue]{hyperref}
\newlength{\extraspace}
\newlength{\extraspaces}
\begingroup
\endgroup

\newcommand{\be}{\begin{equation}
\addtolength{\abovedisplayskip}{\extraspaces}
\addtolength{\belowdisplayskip}{\extraspaces}
\addtolength{\abovedisplayshortskip}{\extraspace}
\addtolength{\belowdisplayshortskip}{\extraspace}}
\newcommand{\ee}{\end{equation}}

\newcommand{\ba}{\begin{eqnarray}
\addtolength{\abovedisplayskip}{\extraspaces}
\addtolength{\belowdisplayskip}{\extraspaces}
\addtolength{\abovedisplayshortskip}{\extraspace}
\addtolength{\belowdisplayshortskip}{\extraspace}}
\newcommand{\ea}{\end{eqnarray}}

\begin{document}

\title{Charged (A)dS black hole solutions in conformal teleparallel equivalent of general relativity}
\preprint{FU-PCG-65}
\author{G.G.L. Nashed$^{1,2,3}$}%
\email{nashed@bue.edu.eg}
\author{Kazuharu Bamba$^{4}$}%
\email{bamba@sss.fukushima-u.ac.jp}
\affiliation{$^{1}$Centre for Theoretical Physics, The British University in Egypt, P.O. Box 43, El Sherouk City, Cairo 11837, Egypt}
\affiliation{$^{2}$Egyptian Relativity Group (ERG), Cairo University, Giza 12613, Egypt}
\affiliation{$^{3}$Mathematics Department, Faculty of Science, Ain Shams University, Cairo 11566, Egypt}
\affiliation{$^{4}$Division of Human Support System, Faculty of Symbiotic Systems Science, Fukushima University, Fukushima 960-1296, Japan}

\begin{abstract}
We continue our study in 4-dimension to derive non-charged and charged (Anti)-de Sitter black hole solutions in conformal teleparallel equivalent of general relativity. The non-charged and charged equations of motion are applied to
a spherically symmetric tetrad and the non-linear differential equations  are derived. It is shown that the output solutions of the two cases are identical to those obtained in teleparallel equivalent theory to general relativity. As a result, it is found that in the conformal teleparallel equivalent theory to general relativity, the scalar field cannot influence on the manifold of  spherical symmetry, i.e. the scalar field must equal 1 in order to have a well known asymptote spacetime.
\vspace{0.2cm}\\
\end{abstract}

\maketitle

\section{Introduction}\label{S1}
To establish a basis of the construction of the Einstein's General Relativity (GR), Deser, Dirac and Utyiama \cite{DESER1970248,Dirac:1973gk,10.1143/PTP.53.565} has introduced the conformal symmetry. The procedure of the conformal invariance to GR has been constructed by Dirac \cite{Dirac:1973gk}. This corresponding to a new variational principle to the action of  GR through the using of a scalar field in addition to the components of the metric $g_{\mu \nu}$ \cite{10.1143/PTP.53.565}. In fact, the Ogievetsky's theory \cite{Ogievetsky:1973ik}, which states  that  {\it  ``GR with diffeomorphism groups can be acquired in the closure of two groups with finite dimension"}, supports to deal with the conformal theory of gravity.

For the quantization of the gravitational field  \cite{tHooft:2011aa}, the conformal symmetry is a significant subject and it is important for the construction of GR with the modifications of the space-time descriptions on the  small or on the large scales.
To establish the formulation of quantum gravity with its renormalizability and unitarity, the amendments of gravitation are necessary on the small scale.
On the other hand, to modify GR on the large scale one can solve the so-called
dark energy problem, namely, realize the late-time cosmic acceleration \cite{Chen:2010va,Capozziello:2011hj,PhysRevD.85.044033}.

For the Einstein-Hilbert action, which is not conformally invariant,
the operation of a conformal transformation can remove
the conformal factor $\Omega(x)$ from the metric tensor $g_{\mu \nu}$
due to the fact that $g_{\mu \nu} =\Omega(x) \hat{g}_{\mu \nu}$.
Hence, it corresponds to an extra degree of freedom and it can be analyzed
as an independent variable \cite{tHooft:2011aa}.
It is known that a scale invariant effective theory can be constructed from $\hat{g}_{\mu \nu}$, which is a non-tensor quantity because of $det \hat{g}=-1$ \cite{Hooft:2010ac}.
The Weyl theory, which is conformally invariant and quadratic in terms of the curvature tensor, has been investigated in various aspects \cite{Mannheim:2005bfa,Moon:2010wq}.

In addition to GR, there is other gravity theory proposed by Einstein, which is called ``Teleparallel Equivalent of General Relativity (TEGR)" \cite{2005physics...3046U}.
In GR, the gravitational field is expressed by the curvature however,
the case of TEGR, is represented by the torsion \cite{osti_4688687,Hayashi:1977jd,PhysRevD.19.3524,Nashed:2003ee,Hanafy:2015yya,PhysRevD.24.3312,Nashed:2019zmy,Blagojevic:1988pp,Kawai:2000vm,
Kawai:2000nj,Maluf1987}. The important advantage to use TEGR is that for the gravity system, energy, momentum and angular momentum can be described in a consistent manner
in the conformal frame \cite{1994JMP....35..335M,Nashed:2011fg,Maluf:2006gu,Maluf:2005kn} (for recent reviews on various proposals for an alternative gravitational theories to GR in order to account for the issue of dark energy, see, for instance,~\cite{Nojiri:2010wj, Capozziello:2011et, Capozziello:2010zz,Awad:2017sau, Bamba:2015uma, Cai:2015emx, Nojiri:2017ncd, Nashed:2015pda,Bamba:2012cp,Nashed:2018qag}).

In terms of the conformal transformation, there exist two descriptions of the Lagrangian for the gravitational field.
One is to introduce an additional scalar field to
the Einstein-Hilbert action. This approach has been taken for the construction of the conformal teleparallel theory in Refs.~\cite{Maluf:2012yn, Silva:2015dea}.
The other is to make the form of the Lagrangian for the gravitational field composed of the quadratic term of the Weyl tensor.

In the context of the conformal teleparallel equivalent of general relativity (CTEGR), it has been demonstrated that the accelerated expansion can be explained in a flat homogenous and isotropic universe \cite{Silva:2015dea}. In this work, we explore the influence of the scalar field on the solar system by applying
the gravitational field equation of the CTEGR to the four-dimensional spherically symmetric space-time.

The organization of the present paper is the following.
In Sec. \ref{S1.1}, we explain the basic formulation of the CTEGR with the definitions of various tensors and the gravitational field equations are presented.
In Sec.~ \ref{S2}, we apply a four-dimensional spherically symmetric
vierbein field with diagonal components to the gravitational field equations of the CTEGR, and we show that the scalar field has no effect due to the acquire of  asymptotically flat, i.e. the scalar field should equal 1. 
In Sec.~ \ref{S3}, we explore a vierbein with its non-diagonal spherically symmetric components in the CTEGR and we find the solutions, whose property is resemble to the case of a vierbein with the diagonal components.
In Sec.~ \ref{S4}, we analyze the charged gravitational field equations of the CTEGR theory
for both the diagonal and non-diagonal vielbeins. These charged gravitational field equations are applied to the vierbein with the diagonal components for the static case. Furthermore, we derive a general charged solution with a physical meaning. In addition, we investigate the charged CTEGR  formalism for the space-time with a non-diagonal components and has a spherical symmetry and derived  solutions in the same way as the case of the space-time with the diagonal components. We show that the only physical solution makes the scalar field has a unit value.
Finally, we summarize our results in Sec.~\ref{S6}.

\section{The conformal teleparallel equivalent of general relativity (CTEGR)}\label{S1.1}
The teleparallel equivalent of general relativity (TEGR) is represented as\footnote{Here, the Latin indices ${\it i, j, \cdots}$ show the coordinates of the tangent space-time, while the Greek ones $\alpha$, $\beta$, $\cdots$ denote the label of the components of the (co-)frame.} $\{{\it M},~L_{i}\}$ with $\it M$ the four-dimensional space-time  manifold and $L_{i}$ ($i=1,2,3,4$) the vectors defined on the space-time $\it M$ globally, which are regarded as the parallel vectors. In the four-dimensional space-time,
the parallel vectors are considered to the {\it vierbein (or tetrad)} fields.
The contravariant derivative of the vierbein (tetrad) field reads
\begin{equation}\label{q1}
  D_{\mu} {L_i}^\nu:=\partial_{\mu}
{L_i}^\nu+{\Gamma^\nu}_{\lambda \mu} {L_i}^\lambda= 0.
\end{equation}
Here, the differentiation is calculated in terms of the Weitzenb\"{o}ck connection, ${\Gamma^\nu}_{\lambda \mu}$, which is the affine connection without symmetry, defined as \cite{Weitzenbock:1923efa}
\begin{equation}\label{q2}
{\it {\Gamma^\lambda}_{\mu \nu} := {L_i}^\lambda~ \partial_\nu L^{i}{_{\mu}}},
\end{equation}
where \[\partial_{\nu}:=\frac{\partial}{\partial x^{\nu}}.\]
The metric is described by
\begin{equation}\label{q3}
 {\rm g_{\mu \nu} :=  \eta_{i\, j} {L^i}_\mu {L^j}_\nu},
\end{equation}
where $\eta_{i j}=(+,-,-,- )$ denotes
the four-dimensional Minkowski space-time.
Equation (\ref{q1}) leads to the condition of the metricity.
We define the torsion tensor ${\mathrm{T}^\alpha}_{\mu \nu}$ and
the contortion one $\mathrm{K}^{\mu \nu}{}_\alpha$ by the following
expressions
\begin{eqnarray} \label{q33}
\nonumber {T^\alpha}_{\mu \nu}  & := &
{\Gamma^\alpha}_{\nu \mu}-{\Gamma^\alpha}_{\mu \nu} ={L_i}^\alpha
\left(\partial_\mu{L^i}_\nu-\partial_\nu{L^i}_\mu\right),\\
{K^{\mu \nu}}_\alpha  & := &
-\frac{1}{2}\left({T^{\mu \nu}}_\alpha-{T^{\nu
\mu}}_\alpha-{T_\alpha}^{\mu \nu}\right). \label{q4}
\end{eqnarray}
By contracting the torsion, we have its vector
\begin{equation}\label{Tv}
{\rm T_\nu := {T^\mu}_{\mu \nu}}.
\end{equation}
Moreover, the torsion scalar in the TEGR is represented as
\begin{equation}\label{Tor_sc}
{\rm T := {T^\alpha}_{\mu \nu} {S_\alpha}^{\mu \nu}}.
\end{equation}
Here, ${S_\alpha}^{\mu \nu}$ is the tensor of the superpotenial.
For the first pair, this tensor has skew symmetry and described by
\begin{equation}\label{q5}
{\rm {S_\alpha}^{\mu \nu} := \frac{1}{2}\left({K^{\mu\nu}}_\alpha+\delta^\mu_\alpha{T^{\beta
\nu}}_\beta-\delta^\nu_\alpha{T^{\beta \mu}}_\beta\right)}.
\end{equation}
The Lagrangian for the TEGR is written as
\begin{equation}\label{q6}
\mathcal{L}({L^i}_\mu)_{g}=\frac{|L|T}{2\kappa}, \qquad \qquad \textrm {where} \qquad \qquad L=\sqrt{-g},
\end{equation}
where $\kappa=8\pi$ is the coupling of the gravitational constant.

The Lagrangian in Eq.~(\ref{q6}) will be changed through the conformal transformation
\begin{equation}\label{q77}
\bar{L}_{a \mu}=e^{\omega(x)} L_{a \mu},
\end{equation}
with $\omega(x)$ being an arbitrary function~\cite{Maluf:2012yn}.
Hence, Eq. (\ref{q6}) should be modified
so that the conformal transformation in Eq.~(\ref{q77}) cannot change
the form of Eq. (\ref{q6}).
The form of Lagrangian that is not changed by the conformal transformation
is known to have the form~\cite{Maluf:2012yn}
\begin{equation}\label{q7}
\mathcal{L}({L^i}_\mu, \Phi)_{g}=2\kappa|L|\left[-\Phi^2T+6g^{\mu \nu}\partial_\mu \Phi\partial_\nu \Phi-4g^{\mu \nu}\Phi(\partial_\nu \Phi) T_\mu\right]+\mathcal{L}_m.
\end{equation}
Here, $\Phi$ is a scalar field and $\mathcal{L}_m$ means the matter-field Lagrangian.
The Lagrangian in Eq.~(\ref{q7}) is invariant under
the transformation for the scalar field
\[ \Phi \rightarrow {\bar \Phi}=e^{-\omega(x)} \Phi.\]
By taking the variation of the Lagrangian (\ref{q7})
in terms of $\Phi$, we get~\cite{Maluf:2012yn}
\begin{equation}\label{q8}
I\equiv \partial_\mu(Lg^{\mu \nu} \partial_\nu \Phi)-\frac{L}{6}\Phi R-\frac{\kappa}{6}\frac{\delta \mathcal{L}_m}{\delta \Phi}=0,
\end{equation}
where the scalar curvature $R$ is represented as
$$LR=2\partial_\nu(LT^\nu)-LT,$$  with $|L|=L=\sqrt{-g}=\det\left({h^a}_\mu\right)$ being the determinant of the metric and $T$ being the scalar torsion.

With the same procedure, the variation of the Lagrangian in Eq.~(\ref{q7})
in terms of the vierbein $L_{a \mu}$ leads~\cite{Maluf:2012yn}
\begin{eqnarray}\label{q9}
&&Q^{a \nu}\equiv \partial_\alpha(L\Phi^2 S^{a \nu \alpha})-L\Phi^2(S^{b  \alpha \nu} T_{b \alpha}{}^a-\frac{1}{4}L^{a \nu} T)
 -\frac{3}{2}LL^{a \nu}g^{\beta \mu}\partial_\beta \Phi \partial_\mu \Phi+3LL^{a \mu}g^{\beta \nu}\partial_\beta \Phi \partial_\mu \Phi
 +LL^{a \nu}g^{\beta \mu}  T_\mu \Phi \partial_\beta \Phi\nonumber\\
& & -L\Phi L^{a \beta}g^{\nu \mu} ( T_\mu  \partial_\beta \Phi+T_\beta  \partial_\mu \Phi)
 -Lg^{\beta \mu}  \Phi T^{\nu a}{}_\mu \partial_\beta \Phi-\partial_\mu[L g^{\beta \nu} L^{a \mu} \Phi \partial_\beta \Phi]+\partial_\rho[L g^{\beta \rho} L^{a \nu} \Phi \partial_\beta \Phi]-\frac{\kappa}{2}\frac{\delta \mathcal{L}_m}{\delta L_{a \nu}}=0,
\end{eqnarray}
where $$\frac{\delta \mathcal{L}_m}{\delta L_{a \nu}}=LL^a{}_\mu {\mathop{\mathfrak{T}}}^{\mu \nu}.$$

For $\Phi=1$, the gravitational field equation (\ref{q9}) is equivalent to
that in the TEGR case. In addition, the trace of Eq. (\ref{q8}) reads
\[\Phi \frac{\delta \mathcal{L}_m}{\delta \Phi}=L\mathop{\mathfrak{T}},\]
where $\mathop{\mathfrak{T}}$ is the trace of the energy-momentum tensor that is
given by
\begin{equation}
\mathop{\mathfrak{T}}=g_{\nu \mu} {\mathop{\mathfrak{T}}}^{\nu \mu}.
\end{equation}
Thus, in the case of the traceless energy-momentum tensor,
$\frac{\delta \mathcal{L}_m}{\delta \Phi}$ is equal to zero.

\section{ Black hole solutions with spherical symmetry: a diagonal tetrad}\label{S2}
The gravitational field equations (\ref{q8}) and (\ref{q9}) of the CTEGR theory
are applied to the four-dimensional tetrad having spherical symmetry and presented in the polar coordinate ($r$, $\theta$, $\phi$, $t$)  by \cite{Nashed:2008ys,Nashed:2015qza,Nashed:2009hn}:
\begin{equation}\label{tetrad}
\hspace{-0.3cm}\begin{tabular}{l}
  $\left({L_{i}}^{\mu}\right)=\left(\frac{1}{\sqrt{N(r)}}, \; r, \; r\sin\theta\; , \sqrt{K(r)}\; \right)$
\end{tabular}
\end{equation}
with $N(r)$ and $K(r)$ are arbitrary functions in terms of the radial coordinate $r$. By combining Eqs. (\ref{tetrad}) and (\ref{q33}),
it is found that the torsion $T^{a b c}$ and contorsion $K^{a b c}$ have the following non-zero components\footnote{The tensor $T^{a b c}$ is represented as \[T^{a b c}=L^a{}_\mu L^b{}_\nu L^c{}_\alpha T^{\mu \nu \alpha}.\] The two tensors $K^{\mu \nu \alpha}$ and $S^{\mu \nu \alpha}$ are defined by the same manner. It is important to stress on the fact that the last two indices of the  superpotential  and the first two ones of the contorsion  are skew symmetry.}
\begin{eqnarray}\label{TC1}
 && T^{(2) (2) (1)}=T^{(3) (3) (1)}=\frac{\sqrt{N}}{r}, \qquad  T^{(4) (1) (4)}=\frac{\sqrt{N}K'}{2K}, \nonumber\\
& & T^{(3) (3) (2)}=\frac{\cot\theta}{r}, \qquad T_{(1)}=-\frac{4K+rK'}{2rK}, \qquad T_{2}=-\cot \theta,\nonumber\\
& & K^{(2) (1) (2)}=K^{(3) (1)  (3)}=\frac{\sqrt{N}}{r}, \qquad  K^{(1) (4) (4) }= \frac{\sqrt{N}K'}{2K},\nonumber\\
& & K^{(3) (2) (3) }= \frac{\cot\theta}{r}.
\end{eqnarray}
Using Eq.  (\ref{TC1}) in Eq. (\ref{q5}),  the superpotential has the following
non-zero components:
\begin{eqnarray}\label{sup}
 && S^{ (1) (2)  (1)}=S^{ (4) (4) (2)}=\frac{\cot\theta}{2r}, \qquad  S^{(4) (4) (1)}= \frac{\sqrt{N}}{r},\nonumber\\
& & S^{(2) (1) (2)  }=S^{(3) (1) (3)  }= \frac{\sqrt{N}(2K+rK')}{4rK}.
\end{eqnarray}
Substituting Eqs. (\ref{TC1}) and (\ref{sup}) into Eq. (\ref{Tor_sc}),
the torsion  and Ricci scalars are described by
\begin{eqnarray}\label{df1}
&&T=\frac{2N(K+rK')}{r^2K},\nonumber\\
&& R=\frac{4rK^2N'-r^2NK'^2+2r^2NKK''+r^2KN'K'+4NK^2-4K^2+4rKNK'}{r^2K^2}.
\end{eqnarray}
By using Eqs. (\ref{TC1}), (\ref{sup}) and (\ref{df1}) in the field equations (\ref{q7}) and (\ref{q8}) with vanishing energy-momentum tensor, i.e.,  ${\mathop{\mathfrak{T}}}^{\nu \mu}=0$,
we obtain
\begin{eqnarray}\label{df11}
& & Q^{(1)}{}_r\equiv 3r^2NK\Phi'^2+4rNK\Phi \Phi'+r^2N\Phi K'\Phi'-K\Phi^2+NK\Phi^2+rN\Phi^2K' =0,\nonumber\\
& &\nonumber\\
& &  Q^{(2)}{}_{\theta}= Q^{(3)}{}_{\phi}\equiv \Phi^2[\{2KN'+2NK'+r N'K'+2rNK''\}K-rNK'^2]+4KN\Phi\Phi'[2K+rK']\nonumber\\
&&\nonumber\\
& &+4rK^2[N'\Phi\Phi'-N\Phi'^2+2N\Phi\Phi'']=0, \nonumber\\
& & \nonumber\\
& &Q^{(4)}{}_t\equiv r\Phi^2N'+N\Phi^2+4rN\Phi\Phi'+r^2N'\Phi\Phi'-r^2N\Phi'^2+2r^2N\Phi\Phi''-\Phi^2=0, \nonumber\\
& &\nonumber\\
& & I\equiv 4rK^2N'\Phi-r^2NK'^2\Phi+2r^2NKK'' \Phi+r^2KN'K'\Phi+4NK^2\Phi-4K^2\Phi+4rNKK'\Phi+6r^2K^2N'\Phi'\nonumber\\
&&\nonumber\\
& &+24rNK^2\Phi'+6r^2NKK'\Phi'+12r^2NK^2\Phi''=0.
\end{eqnarray}

Clearly, it is seen from Eq. (\ref{df11}) that the values of the scalar field $\Phi$ and the two unknown functions $K$ and $N$ cannot be fixed. Therefore, we assume that $\Phi$ has an arbitrary value and discuss all of its possible values, which can be taken physically\footnote{The reason why the unknown functions $K$, $N$ and $\Phi$ cannot be determined by the system of differential equations in (\ref{df11}) is that there exist four differential equations, while we have three unknowns functions.
Only if the scalar field is arbitrary, the system (\ref{df11}) can be reduced to two differential equations in terms of two unknown functions. Any other assumption cannot make the number of the differential equations equal to that of unknown functions. The consequence that the field equations in the CTEGR theory are not able to fix all the unknown functions is valid through the whole of the present study.}.
Hence, the exact solution of Eq. (\ref{df11}) with arbitrary value of the scalar field is represented as
\begin{eqnarray}
& &  \Phi=\Phi,\qquad \qquad  K(r)=\frac{1}{\Phi^2}+\frac{c_1}{r\Phi^3}, \qquad \qquad N(r)=\frac{(r\Phi+c_1)\Phi(r)}{r(\Phi+r\Phi')^2}.
\end{eqnarray}
Here, ``$'$'' denotes the differentiation w.r.t. the radial coordinate $r$ and
$c_1$ is an integration constant.

As a general expression, we analyze the form $\Phi(r)=\frac{1}{r^n}$, where $n$ is a positive integer, i.e., $n\geq0$\footnote{The form of $n<0$ is not allowed since it makes the potentials metric, $K$ and $N$, have no finite value}. In this case, the unknown functions $N(r)$ and $K(r)$ take the following forms
\begin{equation}\label{rr}
K(r)=r^{2n}+c_1r^{3n-1}, \qquad \qquad N(r)=1+c_1 r^{n-1}.
\end{equation}
By using Eq. (\ref{q3}), we obtain the line-element of the solution in (\ref{rr}) as
\begin{equation}\label{rr1}
ds^2=K(r)dt^2-N^{-1}(r)dr^2-r^2(d\theta^2+\sin^2\theta d\phi^2)=(1+c_1 r^{n-1})dt^2-(r^{2n}+c_1r^{3n-1})dr^2-r^2(d\theta^2+\sin^2\theta d\phi^2).
\end{equation}
Equation (\ref{rr1}) shows in a clear way that the line-element corresponding to the scalar field $\Phi(r)=\frac{1}{r^n}$ is meaningless when $n>0$ because it has no well-known\footnote{Physical solution means that a solution behaves as a flat space-time or (Anti-)de Sitter asymptotically.} asymptote behavior in the limit $r\rightarrow \infty$. When $n=0$ the scalar field $\Phi$ is given by\footnote{From Eq. (\ref{sol1}), it can be seen that we have two redundant differential equations of the system (\ref{df11}). The responsible field of this redundancy is the scalar field as we discussed above.}
\begin{eqnarray} \label{sol1}
& &  \Phi(r)=1,\qquad \qquad  N(r)=K(r)=1+\frac{c_1}{r}.
\end{eqnarray}
For Eq. (\ref{sol1}), the metric is represented as
\begin{eqnarray} \label{met2}
ds^2=\left(1+\frac{c_1}{r}\right)dt^2-\left(1+\frac{c_1}{r}\right)^{-1}dr^2-r^2(d\theta^2+\sin^2\theta d\phi^2).
\end{eqnarray}

When $c_1=-2m$ with $m$ being the gravitational mass \cite{Nashed:2004pn}, the metric  (\ref{met2}) is asymptotically flat and becomes the Schwarzschild space-time. As a result, it is demonstrated that in the CTEGR,
for vacuum case, if the metric is diagonal and static and it has spherical symmetry, there exists only the Schwarzschild solution, which is also the solution
in the TEGR.
In the next section, we examine whether this consequence comes from the form of the tetrad in Eq. (\ref{tetrad}) or not. For this purpose, we take the non-diagonal tetrad with its spherical symmetry and explore the possibility of
the scalar field $\Phi$ either constant or not.

\section{Black hole solutions  with spherical symmetry: a non-diagonal tetrad}\label{S3}

The gravitational field equations (\ref{q7}) and (\ref{q8}) in the CTEGR
are applied to a static space-time in $4$-dimensions with its non-diagonal
components and spherical symmetry. This space-time is described by
the polar coordinate ($r$, $\theta$, $\phi$, $t$) as \cite{Nashed:uja,2013PhRvD..88j4034N}
\begin{eqnarray}\label{tetrad1}
\nonumber \left({L^{i}}_{\mu}\right)=
\left( \begin{array}{cccc}
      \displaystyle \frac{\sin{\theta} \cos{\phi}}{\sqrt{N(r)}} & r\cos{\theta} \cos{\phi} & -r\sin\theta \sin{\phi}&0\\[7pt]
   \displaystyle\frac{\sin{\theta} \sin{\phi}}{\sqrt{N(r)}}& r\cos{\theta} \sin{\phi} & r\sin\theta \cos{\phi}&0\\[7pt]
   \displaystyle\frac{\cos{\theta}}{\sqrt{N(r)}}& -r\sin{\theta} & 0&0\\[7pt]
  0 & 0 & 0 &   \sqrt{K(r)} \\
  \end{array}
\right).\\
\end{eqnarray}
It is important to emphasize the fact that tetrad in Eq. (\ref{tetrad1}) is not the most general tetrad that possesses spherical symmetry.
The tetrad in  Eq. (\ref{tetrad1}) can be reproduced from the diagonal tetrad in (\ref{tetrad}) and a rotation SO(3) matrix, i.e.,
 \begin{eqnarray}\label{tetrad5}
 &&\small{
\left( \begin{array}{cccc}
      \displaystyle \frac{\sin{\theta} \cos{\phi}}{\sqrt{N(r)}} & r\cos{\theta} \cos{\phi} & -r\sin\theta \sin{\phi}&0\\[7pt]
   \displaystyle\frac{\sin{\theta} \sin{\phi}}{\sqrt{N(r)}}& r\cos{\theta} \sin{\phi} & r\sin\theta \cos{\phi}&0\\[7pt]
   \displaystyle\frac{\cos{\theta}}{\sqrt{N(r)}}& -r\sin{\theta} & 0&0\\[7pt]
  0 & 0 & 0 &   \sqrt{K(r)} \\
  \end{array}
\right)=\left( \begin{array}{cccc}
       \sin{\theta} \cos{\phi} & \cos{\theta} \cos{\phi} & -\sin{\phi}&0\\[7pt]
   \sin{\theta} \sin{\phi}& \cos{\theta} \sin{\phi} & \cos{\phi}&0\\[7pt]
  \cos{\theta}& -\sin{\theta} & 0&0\\[7pt]
  0 & 0 & 0 &   1 \\
  \end{array}
\right)\times\left( \begin{array}{cccc}
      \displaystyle \frac{1}{\sqrt{N(r)}} & 0 & 0&0\\[7pt]
   0& r &0&0\\[7pt]
   0& 0 & r\sin{\theta}&0\\[7pt]
  0 & 0 & 0 &   \sqrt{K(r)} \\
  \end{array}
\right)}.\nonumber\\
&&
\end{eqnarray}
Similarly to the investigations to derive Eq. (\ref{TC1}) in Sec.~III
for the case of the diagonal tetrad, with Eqs. (\ref{tetrad1}) and (\ref{q33}),  it is seen that the torsion $T^{a b c}$ and contorsion $K^{a b c}$
have the following non-zero components\footnote{$T^{a b c}$, $K^{\mu \nu \alpha}$ and $S^{\mu \nu \alpha}$ are written in the same manner as the one in in Sec.~III for the diagonal tetrad.}
\begin{eqnarray}\label{TC2}
 && T^{(1) (2) (1)}=T^{(3) (2) (3)}=\frac{\sin\theta \sin\phi(1-\sqrt{N})}{r},\qquad T^{(1) (3) (1)}=T^{(2) (3) (2)}=\frac{\cos\theta(1-\sqrt{N})}{r},\nonumber\\
&& T^{(2) (1) (2)}= T^{(3) (1) (3)}=\frac{\sin\theta \cos\phi(1-\sqrt{N})}{r}, \qquad  T^{(4) (1) (4)}=\frac{\sin\theta \cos\phi\sqrt{N}K'}{2K}, \nonumber\\
 &&T^{(4) (2) (4)}=\frac{\sin\theta \sin\phi\sqrt{N}K'}{2K},\qquad T^{(4) (3) (4)}=\frac{\cos\theta \sqrt{N}K'}{2K}, \qquad   T_{(1)}=\frac{4K(1-\sqrt{N})-r\sqrt{N}K'}{2r\sqrt{N}K},\nonumber\\
& & K^{(2) (1) (1)}=K^{(2) (3) (3)}=\frac{\sin\theta \sin\phi(1-\sqrt{N})}{r},\qquad K^{ (3) (1) (1) }=K^{ (3) (2) (2) }=\frac{\cos\theta (1-\sqrt{N})}{r},\nonumber\\
& & K^{(1) (2)  (2)}=K^{(1) (3)  (3)}=\frac{\sin\theta \cos\phi(1-\sqrt{N})}{r},\qquad K^{(1) (4) (4)}=\frac{\sin\theta \cos\phi\sqrt{N}K'}{2K}, \qquad  K^{(2) (4) (4) }=\frac{\sin\theta \sin\phi\sqrt{N}K'}{2K},\nonumber\\
& & K^{(3) (4) (4) }=\frac{\cos\theta \sqrt{N}K'}{2K}.
\end{eqnarray}
It follows from Eq. (\ref{TC2}) that the superpotential has the non-zero components
\begin{eqnarray}\label{sup1}
 && S^{ (1) (1)  (2)}=S^{ (3) (3) (2)}=\frac{\sin\theta\sin\phi(2K(1-\sqrt{N})+r\sqrt{N}K')}{4rK},\nonumber\\
& &  S^{ (2) (2) (1)}= S^{ (3) (3) (1)}=\frac{\sin\theta\cos\phi(2K(1-\sqrt{N})+r\sqrt{N}K')}{4rK},\nonumber\\
& & S^{ (2) (2)  (3)}=S^{ (1) (1)  (3)}=\frac{\cos\theta(2K(1-\sqrt{N})+r\sqrt{N}K')}{4rK}, \qquad S^{(4) (1) (4)  }=\frac{\sin\theta\cos\phi(K(1-\sqrt{N}))}{r}, \nonumber\\
& &   S^{(4) (2) (4)  }=\frac{\sin\theta\sin\phi(K(1-\sqrt{N}))}{r}, \qquad S^{(4) (3) (4)  }=\frac{\cos\theta(1-\sqrt{N})}{r}.
\end{eqnarray}
By combining Eq. (\ref{tetrad}) and Eq. (\ref{Tor_sc}),
the torsion scalar is described as\footnote{
The scalar curvature for (\ref{tetrad1}) is equivalent to the second expression in (\ref{df1}).}.
\begin{equation}\label{df2}
T=\frac{2(1-\sqrt{N})(2K(1-\sqrt{N})+r\sqrt{N}K')}{r^2K}.
\end{equation}
By applying Eq. (\ref{tetrad1}) to Eq. (\ref{q8}) for ${{{\cal
T}^{{}^{{}^{^{}{\!\!\!\!\scriptstyle{em}}}}}}}^\nu_\mu=0$,
the non-zero components, which are equivalent to those for the diagonal tetrad
in Sec.~III, are obtained.
Thus, the same procedure developed for the diagonal tetrad can be
used to the non-diagonal one.

As a consequence from the above investigations,
the scalar field $\Phi$ does not influence of the physics
in vacuum and the space-time with its spherical symmetry
for both diagonal and non-diagonal tetrads.
Hence, in the next sections, we explore the non-vacuum space-time.

\section{Solutions of the charged black hole with its spherical symmetry}\label{S4}

In this section, we study the solutions of the charged black hole that has a spherical symmetry.

\subsection{For diagonal tetrad}

In the CTEGR, the Lagrangian, from which the charged gravitational field equation is derived, is given by
\begin{equation}\label{q17}
\mathcal{L}({L^i}_\mu, \Phi)_{g}=2\kappa|L|\left[-\Phi^2T+6g^{\mu \nu}\partial_\mu \Phi\partial_\nu \Phi-4g^{\mu \nu}\Phi(\partial_\nu \Phi) T_\mu\right]+\mathcal{L}_{em}.
\end{equation}
Here, $\mathcal{L}_{em}=-\frac{1}{2}{ F}\wedge ^{\star}{F}$ with $F = dA$
is  the Maxwell fields and $A=A_{\mu}dx^\mu$ is the 1-form
of the gauge (electromagnetic) potential, and $\Phi$ is a scalar filed.
Based on the variation principle, by varying this Lagrangian (\ref{q17})
in terms of $\Phi$, we acquire
\begin{equation}\label{q88}
I\equiv \partial_\mu(Lg^{\mu \nu} \partial_\nu \Phi)-\frac{L}{6}\Phi R=0.
\end{equation}
This field equation is equivalent to Eq. (\ref{q8}) without charge, which is derived from $\frac{\delta \mathcal{L}_m}{\delta \Phi}=0$,
because the trace of the energy-momentum tensor for the Maxwell fields becomes zero.
Moreover, by taking the variation principle for the Lagrangian (\ref{q17})
in terms of the vierbein $L_{a \mu}$, we have \cite{Maluf:2012yn}
\begin{eqnarray}\label{q19}
&&Q^{a \nu}\equiv \partial_\alpha(L\Phi^2 S^{a \nu \alpha})-L\Phi^2(S^{b  \alpha \nu} T_{b \alpha}{}^a-\frac{1}{4}L^{a \nu} T)
 -\frac{3}{2}LL^{a \nu}g^{\beta \mu}\partial_\beta \Phi \partial_\mu \Phi+3LL^{a \mu}g^{\beta \nu}\partial_\beta \Phi \partial_\mu \Phi
 +LL^{a \nu}g^{\beta \mu}  T_\mu \Phi \partial_\beta \Phi\nonumber\\
& & -L\Phi L^{a \beta}g^{\nu \mu} ( T_\mu  \partial_\beta \Phi+T_\beta  \partial_\mu \Phi)
 -Lg^{\beta \mu}  \Phi T^{\nu a}{}_\mu \partial_\beta \Phi-\partial_\mu[L g^{\beta \nu} L^{a \mu} \Phi \partial_\beta \Phi]+\partial_\rho[L g^{\beta \rho} L^{a \nu} \Phi \partial_\beta \Phi]\nonumber\\
& &-\frac{\kappa \Phi^2}{2}(L^{a \rho}F_{\rho \alpha}F^{\nu \alpha}-\frac{1}{4}L^{a \nu}  F_{\alpha \beta}F^{\alpha \beta})=0.
\end{eqnarray}
Furthermore, from the variation of the Lagrangian (\ref{q17}) w.r.t. the gauge potential $A$, we find
\begin{equation} \label{q29}
E^{\mu}\equiv {\rm \partial_\nu \left( \sqrt{-g} \Phi F^{\mu \nu} \right)}=0.
\end{equation}
It is noted that the derived field equations (\ref{q88}), (\ref{q19}) and (\ref{q29}) are equivalent to those for the Maxwell-TEGR theory when $\Phi=1$ \cite{Nashed:2004pn,2013PhRvD..88j4034N}.

We apply Eqs. (\ref{q88}), (\ref{q19}) and (\ref{q29}) to the space-time (\ref{tetrad}) using Eqs. (\ref{TC1}), (\ref{sup}) and (\ref{df1}).
In this case, we also use the representation of the gauge potential, given by
\begin{equation}\label{df3} A = h(r) dt.
\end{equation}
Using all the above information we obtain the non-zero components of the field equations of CTEGR as
\begin{eqnarray} \label{df4}
& & Q^{(1)}{}_r=3r^2NK\Phi'^2+4rNK\Phi \Phi'+r^2N\Phi K'\Phi'-K\Phi^2+NK\Phi^2+rN\Phi^2K'-4\pi r^2Nh'^2\Phi^2 =0,\nonumber\\
& &  \nonumber\\
& & Q^{(2)}{}_{\theta}= Q^{(3)}{}_{\phi}=\Phi^2[\{2KN'+2NK'+r N'K'+2rNK''\}K-rNK'^2]+4KN\Phi\Phi'[2K+rK']\nonumber\\
&&\nonumber\\
& & +4rK^2[N'\Phi\Phi'-N\Phi'^2+2N\Phi\Phi'']+16\pi rNK h'^2\Phi^2=0, \nonumber\\
& & \nonumber\\
& & Q^{(4)}{}_t=rK\Phi^2N'+NK\Phi^2+4rNK\Phi\Phi'+r^2KN'\Phi\Phi'-r^2KN\Phi'^2+2r^2KN\Phi\Phi''-K\Phi^2-4\pi r^2N h'^2\Phi^2=0,\nonumber\\
& & \nonumber\\
& & I= 4rK^2N'\Phi-r^2NK'^2\Phi+2r^2NKK'' \Phi+r^2KN'K'\Phi+4NK^2\Phi-4K^2\Phi+4rNKK'\Phi+6r^2K^2N'\Phi'\nonumber\\
&&\nonumber\\
& & +24rNK^2\Phi'+6r^2NKK'\Phi'+12r^2NK^2\Phi''=0,\nonumber\\
& & \nonumber\\
& & E^t\equiv 2rNKh' \Phi'-rNh'K'\Phi+rKN'h'\Phi+2rNKh'' \Phi+4NKh'\Phi=0,
\end{eqnarray}
with $h'=\frac{dh}{dr}$.
Thus the  system of differential equations (\ref{df4}) is equivalent to the system  (\ref{df11})
for $h(r)=1$. In addition, the system in (\ref{df4}) becomes the one for the Maxwell-TEGR theory with $\Phi=1$ \cite{Nashed:2004pn}.

The general solution of the system of differential equation (\ref{df4}) is described as
\begin{eqnarray} \label{df6}
& &  \Phi(r)=\Phi(r), \qquad \qquad h(r)=c_2+c_3\int \frac{(r\Phi'+\Phi)}{r^2\Phi^3}dr,\nonumber\\
& & N(r)=\frac{c_4}{r^4\Phi'^2+2r^3\Phi\Phi'+r^2\Phi^2}\int\frac{r^4\Phi'^2+2r^3\Phi\Phi'+r^2\Phi^2}{r^2(r\Phi'+\Phi)}dr
+\frac{c_5}{r^4\Phi'^2+2r^3\Phi\Phi'+r^2\Phi^2}\nonumber\\
& &-2\frac{\int[(r\Phi'+\Phi)(\int r\Phi'dr+\int \Phi dr)]dr-(\int r\Phi'dr+\int \Phi dr)\int (r\Phi'+\Phi)dr}{r^2(r\Phi'+\Phi)^2},\nonumber\\
& &
K(r)=-\frac{\int[2(r\Phi'+\Phi)(\int r\Phi'dr+\int \Phi dr)]dr-(2\int r\Phi'dr+2\int \Phi dr+c_3)\int (r\Phi'+\Phi)dr+c_5}{r^2\Phi^4[2\int( r\Phi'+\Phi)(\int r\Phi'dr+\int \Phi dr)dr+(2r\Phi -2\int r\Phi'dr-2 \int\Phi dr-c_4)\int(r\Phi'+\Phi)dr+\Phi^2r^2-r\Phi c_3+c_5]}.\nonumber\\
& &
\end{eqnarray}

Equation (\ref{df6}) shows that explicit form of the scalar field cannot be given.
The reason of this fact can also be discussed similarly to the neutral case.
The different point from the charged case is the fact that we have five differential equations, while there exist four unknown quantities, $\Phi$, $N$, $K$ and the potential $h$.
The only choice that makes the number of differential equations equivalent to that of unknown quantities is to leave the scalar field arbitrary.

We explore all of the possible values that the scalar field can take. For this end, we take $\Phi$ as previous, i.e., 
\be \label{df7}
\Phi(r)=\frac{1}{r^n}.
\ee
 As a consequence for $n>0$, we find
\be \label{df77}
 h(r)=c_2+c_3r^{n-1},\qquad \qquad  N(r)=\frac{1+r^{n-1}[c_4+c_5r^{n-1}]}{(n-1)^2},
\qquad \qquad K(r)=r^{2n}+c_4r^{3n-1}+c_5r^{2(2n-1)},
\ee
where we have put $c_5=-4c_3{}^2$. The line-element of the constructed from Eq. (\ref{df7}) is undefined in the limit $r\rightarrow \infty$. Thus, the choice $\Phi(r)=\frac{1}{r^n}$ is not a physical one similar to the neutral case.

As another form, we choose
\be \label{df8} \Phi(r)=r^s, \qquad s>0.\ee
The substitution of it into Eq. (\ref{df6}) yields
\be \label{df777}
 h(r)=c_2+\frac{c_3}{r^{s-1}},\qquad \qquad  N(r)=\frac{1+r^{-s-1}[c_4+c_5r^{-s-1}]}{(s+1)^2},
\qquad \qquad K(r)=r^{-2s}+c_4r^{-3s-1}+c_5r^{-2(2s+1)}.
\ee
The line-element of the constructed from in Eq. (\ref{df777})
becomes singular in the limit $r\rightarrow \infty$ because of $K=0$.
Thus, the choice $\Phi(r)=r^s$ must be excluded from our considerations.

As one more form, we investigate the expression $\Phi=1$, either $n=0$ or $s=0$.
For it. we acquire
 \be \label{sol5}
\Phi(r)=1, \qquad \qquad q(r)=c_2-\frac{c_3}{r}, \qquad \qquad N(r)=K(r)=1+\frac{c_4}{r}+\frac{c_3}{r^2}.
\ee
In this case, the metric for the solution (\ref{sol5}) is written as
\be \label{met4}
ds^2=\left(1+\frac{c_4}{r}+\frac{c_3}{r^2}\right)dt^2-\left(1+\frac{c_4}{r}+\frac{c_3}{r^2}\right)^{-1}dr^2-r^2(d\theta^2+\sin^2\theta d\phi^2).\ee
With the discussions in Sec.~III, it can be seen that
the metric in Eq. (\ref{met4}) is asymptotically flat, and that
it is reduced to the Reissner-Nordstr\"{o}m space-time if the constant $c_4=-2m$ and $c_3=q$ \cite{Nashed:2004pn}.

\subsection{For non-diagonal tetrad}
Next, we examine the case of the non-diagonal tetrad.
Similarly to the previous subsection,
we combine the field equations. (\ref{q88}), (\ref{q19}) and (\ref{q29})
and the space-time of Eq. (\ref{tetrad1}) in 4-dimensions.
We also use Eqs. (\ref{TC2}), (\ref{sup1}), (\ref{df2}) and the expression of
the gauge potential in Eq. (\ref{df3}).
As a result, we obtain the same differential equations presented in Eq. (\ref{df4})which corresponding to the case of the diagonal tetrad.
Thus, the same investigations and consequences as those for the diagonal tetrad can be found in the case of the non-diagonal tetrad.

\section{Conclusions}\label{S6}
In the present work, we have explored the issue to analyze the solutions of the black hole in the framework of  ``CTEGR''. We have applied the gravitational field equations of the CTEGR in the vacuum case to a diagonal space-time having a  spherical symmetry. We have described the resultant system of differential equations and derived  the analytic solution of this system. Consequently, we have shown that the only solution with a physical meaning, namely, the explicit representation of the specific solution can be acquired for the scalar field is equal to unity, i.e., $\Phi(r)=1$. That is, for the space-time with the spherical symmetry, the black hole solution of CTEGR theory is equivalent to the TEGR one.

In addition, we have investigated the case of the non-diagonal tetrad for the space-time with the spherical symmetry in vacuum. Based on the same procedures as in the case of the diagonal tetrad, we have written the gravitational field equations. We have found the set of the solutions for this system of differential equation, which
is equivalent to that in the case of the diagonal tetrad.
As a result, it has been understood that in the space-time with the spherical symmetry for both the diagonal and non-diagonal tetrads, the vacuum solution of the black hole in the CTEGR is equivalent to that in the TEGR.

Furthermore, we have studied the black hole solution in the non-vacuum case
by deriving the charged gravitational field equations in the CTEGR.
We have adopted these equations to the 4-dimensional space-time with the
spherical symmetry in the case of the diagonal tetrad.
We have obtained the system of differential equation  and derived the set of the solutions for this equation system.
Similarly to the case in vacuum, it has been demonstrated that the
solution with a physical meaning, i.e., the explicit expression of the specific solution is only the Ressiner-Nordstr\"{o} space-time for $\Phi(r)=1$.
It has also been confirmed that with the same procedures to the case of
the non-diagonal tetrad as those for that of the diagonal one,
if the charged gravitational field equations in the CTEGR are applied to
the non-diagonal tetrad, the same consequences as those in the case of the diagonal tetrad are found.

As a result, it has been concluded that the only physical solution of the black hole with its spherical symmetry in the CTEGR with or without the charge, i.e., the Maxwell field, is equivalent to the solution of the black hole
for $\Phi(r)=1$ in the TEGR. From this result, it can be seen that
a scalar filed in the CTEGR does not influence on the physics in
the space-time with the spherical symmetry due to the fact that $\Phi=1$ to have a well know asymptote behavior. In our viewpoint, the main reason of the corner-stone result of this study may be due to the fact that
we have not used the general form of the tetrad that possesses spherically symmetric or may be due to the structure of the field equations of
the CTEGR theory.
The meaning of the structure of the field equations of the CTEGR theory is that this theory contains the field equations derived from the variation of the Lagrangian w.r.t. the tetrad plus its trace, which is not trivial.
This is similarly to the case of $f(R)$ gravity, whose field equations are the ones that comes from the variation from the Lagrangian w.r.t. the metric plus the trace of this equation. As far as we know, till now there is no black hole solution in $f(R)$ that deviates from GR. This viewpoint needs more investigation that will be done elsewhere.

\subsection*{Acknowledgments}
The work of KB was partially supported by the JSPS KAKENHI Grant Number JP
25800136 and Competitive Research Funds for Fukushima University Faculty
(19RI017).
%

\end{document}